# Towards smooth (010) β-Ga$_2$O$_3$ films homoepitaxially grown by plasma assisted molecular beam epitaxy: The impact of substrate offcut and metal-to-oxygen flux ratio

P. Mazzolini[1,a)] and O. Bierwagen[1]

[1] *Paul-Drude-Institut für Festkörperelektronik, Leibniz-Institut im Forschungsverbund Berlin e.V., Hausvogteiplatz 5–7, 10117 Berlin, Germany*

a) *mazzolini@pdi-berlin.de*

Smooth interfaces and surfaces are beneficial for most (opto)electronic devices that are based on thin films and their heterostructures. For example, smoother interfaces in (010) β-Ga$_2$O$_3$/(Al$_x$Ga$_{1-x}$)$_2$O$_3$ heterostructures, whose roughness is ruled by that of the β-Ga$_2$O$_3$ layer, can enable higher mobility 2-dimensional electron gases by reducing interface roughness scattering. To this end we experimentally prove that a substrate offcut along the [001] direction allows to obtain smooth β-Ga$_2$O$_3$ layers in (010)-homoepitaxy under metal-rich deposition conditions. Applying In-mediated metal-exchange catalysis (MEXCAT) in molecular beam epitaxy at high substrate temperatures (T$_g$ = 900 °C) we compare the morphology of layers grown on (010)-oriented substrates having different unintentional offcuts. The layer roughness is generally ruled by *(i)* the presence of (110)- and ($\bar{1}$10)-facets visible as elongated features along the [001] direction (rms < 0.5 nm), and *(ii)* the presence of trenches (5-10 nm deep) orthogonal to [001]. We show that an unintentional substrate offcut of only ≈ 0.1° almost oriented along the [001] direction suppresses these trenches resulting in a smooth morphology with a roughness exclusively determined by the facets, *i.e.,* rms ≈ 0.2 nm. Since we found the facet-and-trench morphology in layer grown by MBE with and without MEXCAT, we propose that the general growth mechanism for (010)-homoepitaxy is ruled by island growth whose coalescence results in the formation of the trenches. The presence of a substrate offcut in the [001] direction can allow for step-flow growth or island nucleation at the step edges, which prevents the formation of trenches. Moreover, we give experimental evidence for a decreasing surface diffusion length or increasing nucleation density on the substrate surface with decreasing metal-to-oxygen flux ratio. Based on our experimental results we can rule-out step bunching as cause of the trench formation as well as a surfactant-effect of indium during MEXCAT.

**THE MANUSCRIPT**

**Introduction**

Gallium oxide in its most thermodynamically stable monoclinic structure β-Ga$_2$O$_3$ has recently been proposed as a promising material for power electronics.[1] The possibility to deposit it on native substrates grown from the melt[2] can allow for the synthesis of high quality thin films. Nonetheless, the growth of β-Ga$_2$O$_3$ is orientation-dependent, and this can affect both the structural quality[3,4] and the growth rate[4,5] of the deposited layers [*e.g.* structural defects and low deposition rate in (100)-oriented layers]. For these reasons, the most widely employed substrate orientation for β-Ga$_2$O$_3$ homoepitaxy has been so far the (010) one as it prevents formation of twin defects and provides a comparably high growth rate in molecular beam epitaxy (MBE). MBE and metal-organic vapor phase epitaxy (MOVPE) have been so far the deposition techniques that provided high quality homoepitaxial β-Ga$_2$O$_3$ layers, also enabling for a fine control of their electrical properties throughout *n*-type extrinsic doping,[6–8] as well as the growth of modulation-doped heterostructures.[9–11] Nonetheless, different synthesis conditions can affect the electrical properties of the deposited layers through the formation of deep level acceptors that could work as electron traps in β-Ga$_2$O$_3$ (*e.g.,* Ga-vacancies), potentially limiting both charge carrier density and mobility.[12,13] Therefore, Ga-rich deposition conditions[12] and high growth temperatures $T_g$[14] have been theoretically predicted to be favorable for the synthesis of β-Ga$_2$O$_3$ layers. Unfortunately, due to the strong desorption of the intermediately formed volatile suboxide Ga$_2$O from the growth surface,[15] the deposition of Ga$_2$O$_3$ under these conditions is challenging even in the case of (010) homoepitaxy.[16] The employment of an additional In-flux as a catalyst during Ga$_2$O$_3$ deposition in MBE, *i.e.,* metal-exchange catalysis (MEXCAT),[17] has been shown to result in large incorporation of the impinging Ga-flux even under synthesis conditions that would not otherwise allow for layer growth (*e.g.,* metal-rich, high $T_g$) with very limited In-incorporation in the deposited layer.[4,16,17] A similar effect has been also demonstrated using Sn as catalyzing element.[18] In particular, for β-Ga$_2$O$_3$ (010)-homoepitaxy we have shown[16] that In-mediated MEXCAT-MBE provides high quality β-Ga$_2$O$_3$ layers with almost full Ga-flux incorporation under metal-rich deposition conditions at $T_g$ = 900 °C. Moreover, we identified that the surface roughness of the (010)-oriented layers is usually dominated by two distinct morphologies,[16] both identifiable by atomic force microscopy (AFM): *(i)* the presence of (110) and ($\bar{1}$10) surface equivalent facets visible as straight lines oriented along the [001] in-plane direction (highlighted in red in Figure 1), and *(ii)* the presence of trenches/grooves visible as irregular features running almost orthogonal to the facets (blue dotted lines in Figure 1). The presence of *(i)* facets is ruled by thermodynamics [*i.e.,* (110) more stable surface with respect to (010) under reducing/metal-rich conditions], but has nonetheless found to have a limited impact on the overall surface roughness of the deposited layers, since for high $T_g$ it is possible to obtain peak-to-valley height of less than 0.5 nm with lateral spacing of ≈ 5-10 nm.[16] Differently, the *(ii)* trenches/grooves are found to be usually 5-10 nm deep with a typical trench-to-trench distance in the order of 300-500 nm[16] and could therefore be problematic for the realization of heterostructures, *e.g.,* by reducing the mobility of 2-dimensional electron gases (2DEGs) at the interface of modulation-doped single[9–11] or double[19] β-(Al$_x$Ga$_{1-x}$)$_2$O$_3$/Ga$_2$O$_3$ structures. The formation of trenches/grooves on the surface of (010)-oriented β-Ga$_2$O$_3$ and β-(Al$_x$Ga$_{1-x}$)$_2$O$_3$ layers has been widely reported (but little commented) in literature for both MBE[6,7,9,16,20] and MOVPE.[8,21,22]

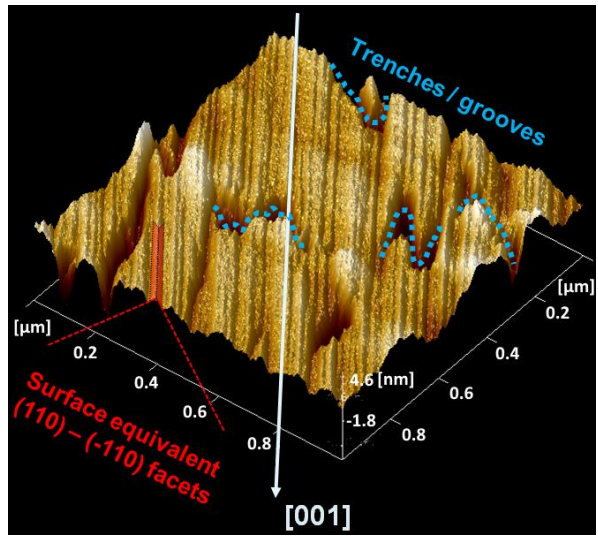

**Figure 1** 3-dimensional representation of an AFM image of an (010) β-Ga$_2$O$_3$ homoepitaxial layer (2d image of the same sample is reported in Figure 3 (c)). The presence of two facets is exemplarily highlighted in red, while some trenches are exemplarily marked as blue dotted lines.

To date, different explanations have been given for the formation of these trenches. Based on homoepitaxial growths by ozone MBE, Sasaki et al.[7] suggested that the groove formation should be related to step-bunching along the [001] direction, given the possibility to reduce the rms of the deposited layers by lowering $T_g$. In contrast, smooth (010)-layers have been obtained by Okumura et al.[6] using plasma-assisted MBE at high substrate temperatures on substrates with a large (2°) unintentional offcut along the [001] in-plane direction, suggesting the absence of step-bunching in favor of step-flow growth. We have recently reported a trench-free (010) β-Ga$_2$O$_3$ homoepitaxial layer deposited at $T_g$ = 900 °C grown by In-mediated MEXCAT via plasma-assisted MBE whereas a layer deposited under the very same conditions (*i.e.* $T_g$ and O-to-Ga flux ratio), but without In-mediated MEXCAT resulted in the formation of trenches, speculating on either an impact of different unintentional offcuts or an increase of the surface diffusion length due to In-mediated MEXCAT.[16] Additionally, it has been shown that In-mediated MEXCAT via plasma-assisted MBE also realized trench-free β-(Al$_x$Ga$_{1-x}$)$_2$O$_3$ layers on (010) β-Ga$_2$O$_3$ substrates,[23] suggesting both a catalytic and a surfactant effect of In. A possible role of In as a surfactant has been also suggested in (100) homoepitaxy of β-(In$_x$Ga$_{1-x}$)$_2$O$_3$ via MOVPE.[24] In this work we experimentally investigate the cause of trench formation in (010) β-Ga$_2$O$_3$ homoepitaxy by plasma-assisted MBE with respect to the substrate offcut and the metal-to-oxygen flux ratio during plasma-assisted MBE.

**Experiment**

We use In-mediated MEXCAT to deposit β-Ga$_2$O$_3$ homoepitaxial layers on top of substrates with different unintentional offcuts previously measured by a combination of X-ray reflectivity (XRR) and X-ray diffraction (XRD - PANalytical X'Pert Pro MRD). An O-plasma treatment at T$_g$ = 900 °C has been always performed prior to the deposition process and resulted in a featureless surface.[16] For the details regarding the experimental process and the substrate characterization please consult ref.[16]. The offcut measurements were collected on 10x15 mm$^2$ substrates which were afterwards cut in 5x5 mm$^2$ pieces. For their in-plane orientation the skew-symmetric reflection of the (111) crystal plane was measured. We performed four different offcut measurements by rotating the in-plane direction (Φ angle) of the sample in steps of 90°, each time aligning ω for the (020) crystal planes and afterwards measuring the shift of the surface reflection with respect to this alignment [2-Theta = 0.5°, 1/16° beam width, 0.18 mm detector slit, see Figure 2*(a)*]. The four offcut measurements were fitted with a sine function in order to determine the absolute offcut component and direction [Figure 2(b)].

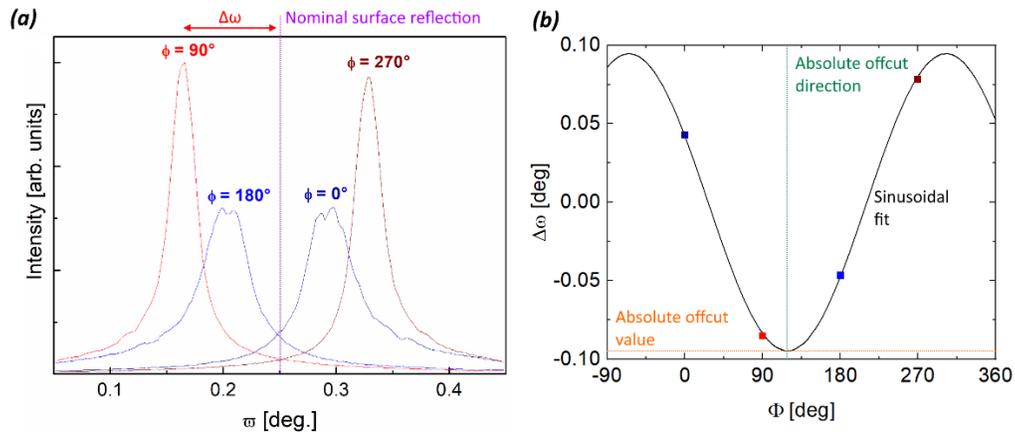

**Figure 2** *(a)* XRD ω scans performed on the 15x10 mm$^2$ unintentionally doped Ga$_2$O$_3$ substrate (see Table 1) in total reflection configuration (2-Theta = 0.5°) in four different in-plane directions (Φ-angles) separated by 90° after ω alignment to the (020) crystal planes. For example, the Φ = 90° measurement (red curve) is performed along the [001] direction; the offcut component along that in-plane direction is reflected by the Δω difference (red arrow) with respect to the nominal surface reflection (ω = 0.25°, violet dotted line). *(b)* Sinusoidal fit of the Δω offcut components evaluated in *(a)* for the same sample for the obtainment of absolute offcut value and direction (orange and green dotted line respectively).

We found that with our experimental setup the measurement of the substrate offcut on (010) substrates can be reliably done just before the full crystal cut in 5x5 mm$^2$ pieces; we believe this could be related to the round edges on the sides of the (010) crystals which could induce a broadening of the surface reflection component in such a small sample size. A sum up of the measured unintentional offcuts α on the 4 different employed substrates is reported in Table 1. The expected associated terrace length has been evaluated considering monolayer steps equal to the *b* unit cell parameter, *i.e.* 0.303 nm;[25] nonetheless, consistently to what we have previously reported,[16] we never identified monolayer steps before the deposition process. The film thickness was determined from XRD "Pendellösung" fringes in the vicinity of the (020) *β*-$Ga_2O_3$ reflection in 2Θ-ω scans.[6,16] The surface morphology of the deposited layers was characterized by AFM (Bruker Dimension Edge) in the PeakForce tapping mode on two different image sizes (1x1 – 5x5 μm$^2$).

| Substrate | Absolute offcut α [°] | Expected terrace length [nm] | Absolute offcut in-plane direction with respect to [001] [°] | Offcut component along [001] [°] |
|---|---|---|---|---|
| Fe:$Ga_2O_3$ - 1 | 0.12 | 150 | 86 | 0.01 |
| Fe:$Ga_2O_3$ - 2 | 0.13 | 135 | 55 | 0.07 |
| Sn:$Ga_2O_3$ | 0.04 | 435 | 41 | 0.02 |
| unintentionally doped $Ga_2O_3$ | 0.09 | 195 | 28 | 0.08 |

**Table 1** Measured unintentional offcuts of the (010) β-$Ga_2O_3$ substrates used in this study. Exemplary measurement of unintentionally doped $Ga_2O_3$ substrate reported in Figure 2.

We deposited a series of samples with MEXCAT at $T_g$ = 900 °C under identical, nominally metal-rich conditions[16] ($\phi_{Ga}$ = 2.2 nm$^{-2}$s$^{-1}$, $\phi_{In}$ = 0.7 nm$^{-2}$s$^{-1}$, O-flow = 0.33 sccm – plasma power P = 300 W, growth time 30 minutes) on top of the four characterized substrates with different unintentional offcuts (Table 1). In line with our previous report,[16] under these growth conditions we were able to obtain for all the deposited layers a comparable thickness (range of 80-100 nm) without XRD-detectable In incorporation [*i.e.,* without shift of the layer peak with respect to the (020) reflection of the substrate – *e.g.,* red curve in Figure 6(d)]. Moreover, the effect of three different oxygen flows (*i.e.*, 0.2 – 0.33 – 0.5 sccm) was investigated while maintaining constant the other deposition parameters for layers deposited on top of substrates obtained from the same Sn:$Ga_2O_3$ crystal (see Table 1).

## **Results**

Figure 3 summarizes the AFM micrographs of all the layers deposited under these conditions. They all show homogeneous morphologies as visible comparing 1x1 and 5x5 μm$^2$ images.

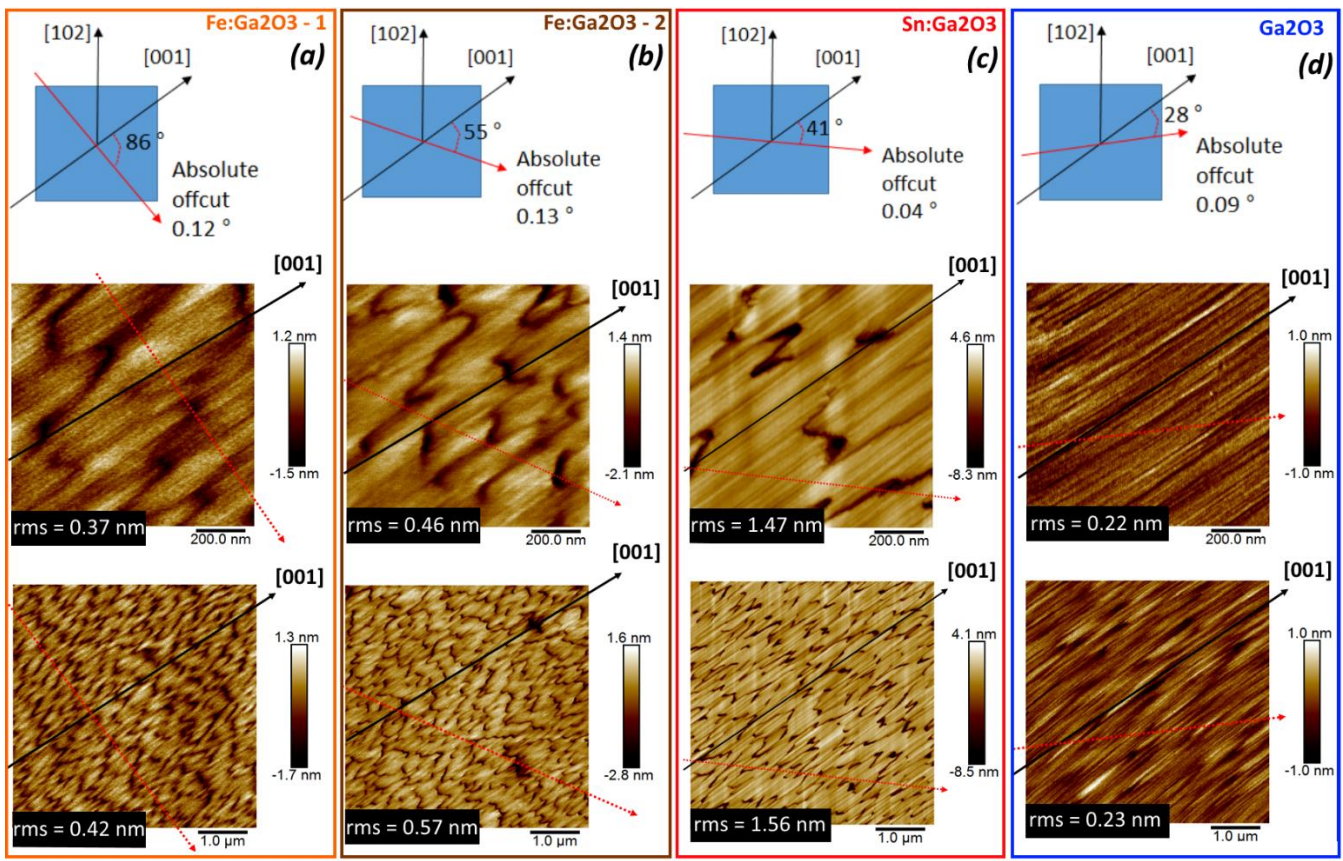

**Figure 3** 1x1 - 5x5 µm² AFM mages of (010) homoepitaxial layers grown under identical conditions by MEXCAT-MBE on four different substrates (a-d). The in-plane orientations and measured absolute offcut directions and absolute angles are marked by black and red arrows respectively in both a representative sketch of the substrates (top view) and the respective acquired AFM images of the layers deposited on top of them.

The samples deposited on substrates having the in-plane direction of their absolute offcuts 86°, 55°, and 41° away from the [001] direction [Figure 3(a), (b), and (c) respectively], all show the discussed morphological features typical of (010)-homoepitaxial layers, *i.e.* (110)-facets visible as straight lines oriented along the [001] direction, and several nanometer-deep irregularly shaped trenches/grooves almost orthogonally oriented to the [001] direction. In order to highlight the trenches, we report in Figure 4 exemplary line profiles extracted along the [001] direction from the reported 5x5 µm² AFM images of Figure 3 (color code of Figure 4 in accordance to Figure 3). In the case of the layers deposited on the two substrates having similar absolute offcuts (0.12° and 0.13°) with in-plane orientation 86° and 55° from the [001], despite a low mean roughness of about 0.5 nm [Figure 3 (a) and (b)], they both show similar trench-to-trench mean distance (≈ 450 nm) and trenches mean height of about 1 – 2 nm as indicated by the yellow and brown lines in Figure 4 (note that the trench depth extracted from our AFM images is a lower bound estimate as it may be limited by a too-large tip radius). Differently, the layer deposited on the substrate having lower in-plane offcut orientation (41°) from the [001] and the lowest absolute value (0.04°) results in a significantly rougher layer [rms ≈ 1.5 nm, Figure 3(c)]. This is related to the formation of deeper trenches (≈ 4 – 6 nm, red line in Figure 4); nonetheless, we notice that the trench-to-trench mean distance is also significantly increased from the ≈ 450 nm of previous samples, to about 750 nm. Notably, the layer deposited on top of the substrate having the closest absolute offcut to the [001] in-plane direction [28°, absolute offcut value 0.09° Figure 2 and Figure 3(d)] is showing a trench-free morphology (blue line in Figure 4).

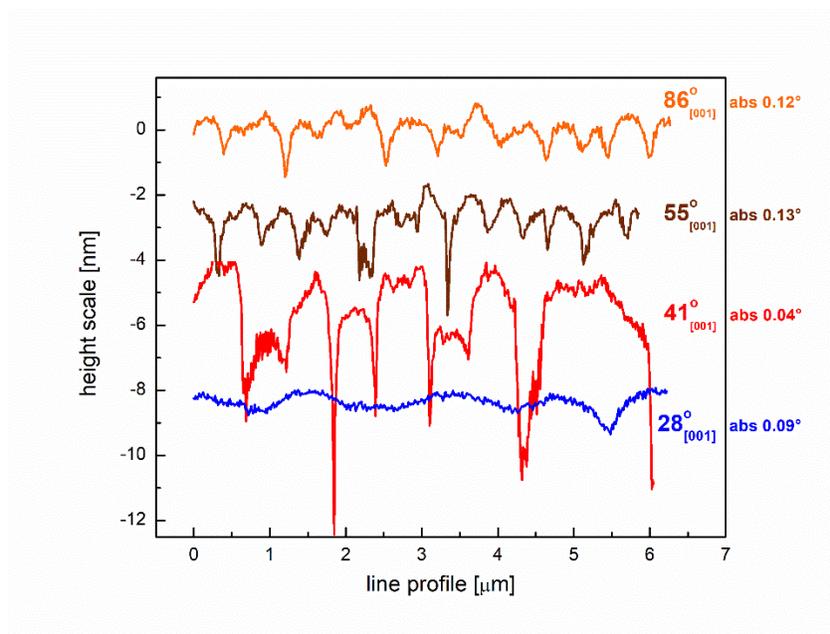

**Figure 4** Line profiles extracted along the [001] direction from the 5x5 µm$^2$ AFM images reported in Figure 3 (line color code).

In order to confirm our results, we performed a twin deposition (same growth conditions) on top of a second 5x5 mm$^2$ unintentionally doped substrate coming from the same 15x10 mm$^2$ crystal (reference deposition Figure 3(d) and blue line in Figure 4). The AFM pictures of the deposited layer [Figure 5(a)] confirm the presence of a layer without the presence of trenches, whose rms is slightly lower than its twin deposited sample (rms ≈ 0.15 nm) and is just ruled by the (110)-faceting. Remarkably, it is possible to identify in this case the presence of monolayer steps whose direction and spacing [Figure 5(a) and (b) respectively] is in line with the measured substrate offcut (Table 1 and Figure 2). This evidence points towards a potential step-flow growth of the deposited (010) homoepitaxial layer.

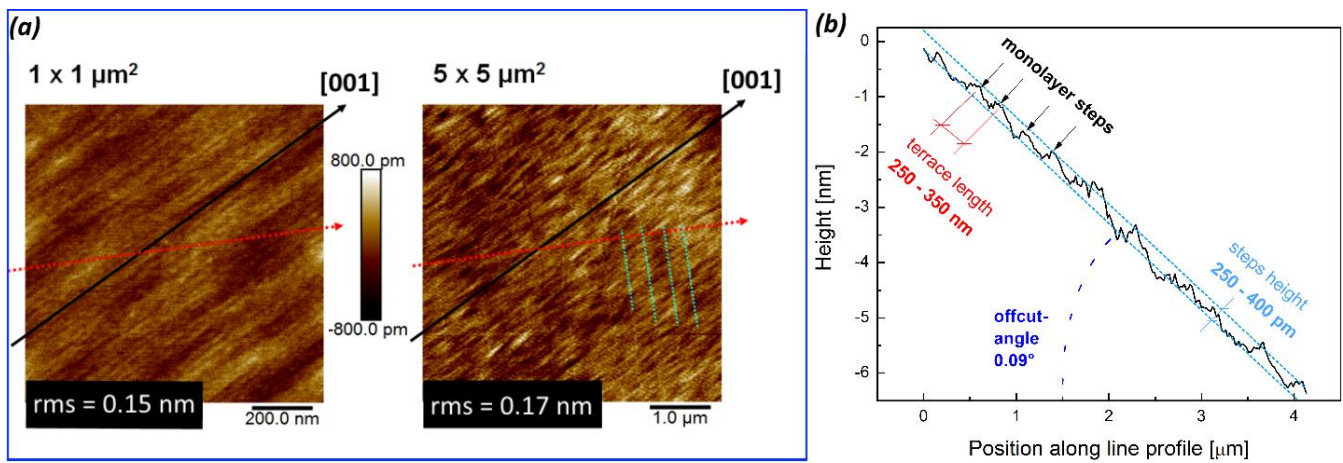

**Figure 5** (a) AFM micrographs of a twin homoepitaxial deposition on top of a second 5x5 mm² piece of the unintentionally doped crystal (reference Figure 3(d)); the absolute offcut direction of the substrate is marked as a red arrow (details in Table 1), while the presence of monolayer steps is evidenced by dotted cyan lines on the 5x5 µm² image. (b) Line profile extracted from the 5x5 µm² AFM image along the offcut direction. A line was added to the profile to correctly reflect the offcut angle.

Finally, we investigated the effect of different Ga-to-O fluxes. We deposited on top of two 5x5 mm² Sn:Ga$_2$O$_3$ substrates from the same investigated 15x10 mm² crystal (Table 1) at slightly higher (0.5 sccm) and lower (0.2 sccm) O-flows [reference O-flow = 0.33 sccm, Figure 3(c)] while maintaining the same O-plasma P, T$_g$, and In-flux during the MEXCAT growth. A 0.2 sccm O-flow results in a surface morphology dominated by rougher (110)-facets with respect to the 0.33 sccm layer [Figure 6(a) and Figure 3(c) respectively]. From Figure 6(a) the presence of trenches is not clearly evident as in the case of the 0.33 sccm layer [Figure 3(c)]; the extracted line profiles along the [001] direction of the 0.2 sccm layer show the presence of smoother valleys with respect to the trenches evidenced in the sample deposited at 0.33 sccm [black and red lines respectively in Figure 6(c)], but similar valley-to-valley periodicity (*i.e.* ≈ 750 nm). The XRD of the sample deposited at 0.2 sccm indicated neither In-incorporation nor thickness fringes [black curve in Figure 6(d)], but it is reasonable to expect a thinner layer with respect to the 0.33 sccm layer [≈ 80 nm from XRD, see thickness fringes in red curve of Figure 6(d)] due to the lower O-flow.[15]

The slightly higher O-flow of 0.5 sccm resulted in the formation of clear trenches [Figure 6(b)] whose mean trench-to-trench distance is strongly reduced with respect to the 0.33 sccm sample [≈ 200 nm vs. ≈ 750 nm, green and red line respectively in Figure 6(c)]. Moreover, the O-richer deposition conditions resulted in *(i)* the presence of less defined (110)-facets [Figure 6(b)] and *(ii)* the incorporation of a detectable amount of In [(020) left side shift, green line in Figure 6(d)]. This is expected given *(i)* the stability of the (010) surface under oxygen richer atmospheres[16] and *(ii)* the tendency to incorporate In in MEXCAT-MBE at high O-fluxes.[4]

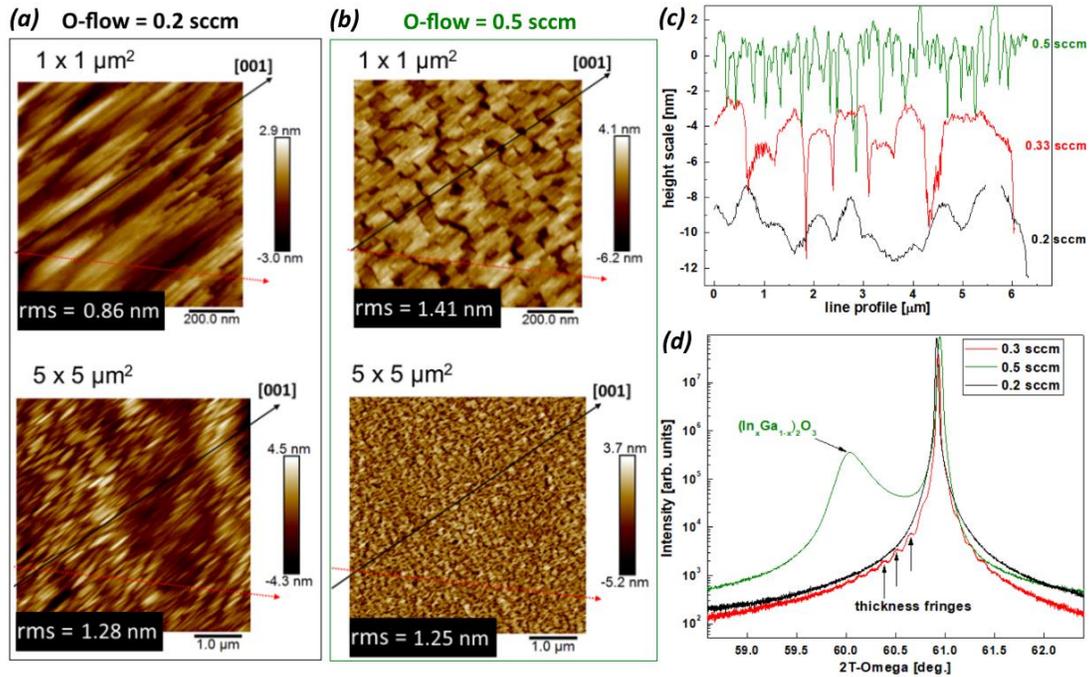

**Figure 6** (a) and (b) AFM micrographs of a homoepitaxial depositions on top of Sn-doped substrate at respectively lower (0.2 sccm) and higher (0.5 sccm) O-flow with respect to the reference layer [0.33 sccm, Figure 3(c)]. (c) Line profiles extracted along the [001] direction from the respective 5x5 µm² AFM images. (d) XRD of the reference layer (red curve) and the ones deposited at O-flows of 0.2 and 0.5 sccm (black and green curves respectively).

## **Discussion**

We interpret the collected experimental results by an anisotropic diffusion lengths and associated nucleation densities of the adsorbed species on the (010) $Ga_2O_3$ surface. This becomes obvious already by considering the morphologies of films, shown in Fig. 4 of Ref.[16] that were grown without MEXCAT but at otherwise identical conditions to those discussed here. In particular, we can consider the case of *(i)* the (110)-facets and *(ii)* the trenches for the layers deposited at slightly metal-rich conditions, *i.e.* 0.33 sccm. The *(i)* facets lateral size of about 5-10 nm[16] is a good indication of the typical diffusion length of the adsorbed species along the orthogonal direction to [001]. On the other hand, the *(ii)* extent of the facets in the [001] direction as well as the large mean distance of about 300-500 nm[16] among trenches are suggesting a much larger diffusion length and associated lower nucleation density along the [001] direction. The experimental data of the present work for films grown under same metal-to-oxygen flux ratio and substrate temperature but with In-mediated MEXCAT does not change this facet-and-trench picture with respect to non-catalyzed MBE growth [compare Figure 3 of the present work to Fig.4 of Ref.[16]]. Differently from what has been proposed elsewhere,[23,24] these experimental data suggests that In does not act as a surfactant for the $Ga_2O_3$ layer growth. Instead, our present work indicates that a sufficiently large absolute offcut mostly oriented along the [001] direction (unintentionally doped $Ga_2O_3$ substrate, see Table 1 and Figure 2) is the key for the formation of a trench-free (010) homoepitaxial layer [Figure 3(d) and Figure 5]. The related monolayer steps (highlighted in Figure 5) with spacing below or equal the diffusion length along the [001] direction of the surface-diffusing species can act as a regular array of nucleation sites. As shown by the presence of off-cut-related monolayer steps on the deposited layer shown in Figure 5, this approach can eventually allow for step-flow growth in (010) homoepitaxy. Therefore, the resulting layers are homogeneously smooth with a low surface roughness (rms ≈ 0.2 nm) just ruled by the formation of the (110)-facets. The obtainment of such a low surface roughness despite the faceting is allowed by the low angle of the (110)-facets with respect to the (010) surface (≈ 14°) and their limited lateral size (≈ 5-10 nm).[16] We propose that the employment of proper offcuts could be fundamental for the growth of the highest surface quality β-$Ga_2O_3$ homoepitaxial layers on all the available orientations since a well-defined substrate offcut has already allowed to achieve step flow growth in (100) β-$Ga_2O_3$ homoepitaxy by MOVPE.[26]

Moreover, our data suggest that a too-low offcut and related long distance of monolayer steps (above the surface diffusion length) in the [001]-direction results in the random nucleation of islands whose coalescence forms the trenches during (010) homoepitaxy by plasma-assisted MBE with or without MEXCAT. An exception to this tentative explanation is the film grown on the substrate Fe:$Ga_2O_3$–2 which exhibits trenches (shown in Figure 3 (c)) despite an offcut component along the [001] (Table 1) that is comparable to that of the trench-free film grown on the unintentionally doped $Ga_2O_3$ substrate. Notwithstanding, this discrepancy suggests the actual absolute offcut direction along [001] to play a key role for obtaining trench-free layers.

In addition, the comparison among layers deposited on the same crystal with different O-to-Ga flux ratios allows us to identify the O-flow as an important parameter to change the diffusion length of the adsorbed species on the layer surface. In particular, while maintaining the same metal flux, a larger O-flow decreases the diffusion length (and increases the nucleation density) along the [001] direction resulting in the formation of closer spaced trenches [Figure 6(b,c)]. We conclude that step bunching is not responsible for the formation of trenches that roughen the (010) layers in plasma assisted MBE since this mechanism – contrary to our observations – should be *(i)* promoted by the presence of a shorter distance of monolayer steps along the (high diffusion length) [001] direction rather than a longer one and *(ii)* suppressed/reduced by a shorter diffusion length due to higher O-fluxes (0.5 sccm instead of 0.33 sccm).[27]

**<u>Conclusion</u>**

In conclusion, we found (010) β-Ga$_2$O$_3$ films homoepitaxially grown under slightly metal-rich conditions by plasma-assisted MBE with and without In-mediated MEXCAT to be significantly roughened due to the formation of trenches whereas the faceting into shallow (110)-($\bar{1}$10) facets elongated along the [001] direction plays a minor role for the total film roughness. The trench formation is likely related to coalescence boundaries due to an island growth regime with significantly higher diffusion length, and thus lower nucleation density, along the [001] direction than perpendicular to it. In agreement with this the metal-to-oxygen flux ratio was identified as an important synthesis parameter that controls the diffusion length of the adsorbed species during layer growth and thus the trench density: A higher oxygen flux results in lower diffusion length and thus a higher density of trenches.

Using In-mediated MEXCAT, we demonstrate, that a sufficiently large substrate offcut oriented along the [001] direction can enable the growth of smooth, trench-free layers, *i.e.,* about 100 nm thick layers with rms roughness as low as 0.2 nm, in (010) β-Ga$_2$O$_3$ homoepitaxy by plasma-assisted MBE. The absence of the trenches is tentatively attributed to the formation of proper monolayer steps whose width is comparable to or lower than the diffusion length of the surface diffusing species in that particular in-plane direction, indicating the possibility of step flow growth. The observed dependency of trench formation on offcut and oxygen flux excludes both *(i)* step bunching as their creation mechanism, as well as *(ii)* the possible role of In as a surfactant in (010) β-Ga$_2$O$_3$ homoepitaxy.

We believe that this work can be fundamental for the realization of high quality interfaces and surfaces in multilayer heterostructures like β-(Al$_x$Ga$_{1-x}$)$_2$O$_3$/Ga$_2$O$_3$ on (010) β-Ga$_2$O$_3$ substrates with increased electron mobility due to reduced surface or interface roughness scattering.

**Acknowledgments:**

We would like to thank Martin Albrecht, Konstantin Lion and Stefano Cecchi for fruitful scientific discussion, Martin Heilmann for critically reading the manuscript, as well as Hans-Peter Schönherr and Carsten Stemmler for technical MBE support. This work was performed in the framework of GraFOx, a Leibniz-Science Campus partially funded by the Leibniz Association.